\documentclass[twocolumn,amsmath,amssymb]{revtex4}

\usepackage{graphicx}
\usepackage{dcolumn}
\usepackage{amsmath,amsthm}
\usepackage{amsfonts}
\usepackage{amssymb}
\usepackage{bbold}
\usepackage{amssymb}
\usepackage{epstopdf}
\usepackage{times}
\usepackage{mathrsfs}
\usepackage{color}
\usepackage{gensymb}
\usepackage{times}
\usepackage{subfigure}
\usepackage[]{graphicx}
\usepackage{amssymb}
\usepackage{amsmath}
\usepackage{bm}
\DeclareGraphicsExtensions{.PNG,.jpg,.pdf,.gif,.eps}
\usepackage{amsmath,amssymb,amsthm,amsfonts,eucal}
\usepackage[english]{babel}

\begin{document}
\title{Mechanotaxis and cell motility }
\author{P. Recho, T. Putelat and L. Truskinovsky}

\affiliation{LMS,  CNRS-UMR  7649,
Ecole Polytechnique, Route de Saclay, 91128 Palaiseau,  France}

\date{\today}
\begin{abstract}
We propose a mechanism of cell motility which is based on contraction and does not require protrusion. The contraction driven translocation of a cell is due to internal flow of the cytoskeleton  generated by molecular motors. Each motor contributes to the stress field and simultaneously undergoes biased random motion in the direction of a higher value of this stress. In this way  active cross-linkers use passive actin network as a medium through which they interact and self-organize. The model exhibits motility initiation pattern similar to the one observed in experiments on keratocytes.

\end{abstract}
\maketitle

Coordinated crawling-induced movements of eukaryotic cells involve spatial and temporal self-organization at the cytoskeletal level. In particular, to achieve a  motile  configuration the cell must first  polarize \cite{Alberts2002}.  While both myosin contraction and actin treadmilling contribute to cell migration, contraction appears to be essential for polarization, moreover, cells may be driven by contraction only \cite{Verkhovsky1999}. The contraction-dominated motility is driven by 'pullers' and can take place even when 'pushers' are disabled  \cite{Simha2002}.

In this Letter we show that the positive feedback mechanism giving rise to symmetry breaking involved in contraction dominated motility can be interpreted as an uphill diffusion driven at the microscale by advection of molecular motors. These motors mechanically propel the actin network by inflicting contraction. In turn, the network  drags the motors amplifying contraction and creating an autocatalytic effect \cite{Mayer2010}. Such coupling leads to build up of motor concentration which is limited by elastic stiffness, friction and diffusion, all resisting the runaway and providing a negative feedback.

By using the term  \emph{mechanotaxis}  we imply conceptual similarity of the described motility mechanism with chemotaxis. Each motor generates a stress field and the other motors undergo biased random motion in the direction of a higher value of the stress. In this way  active cross-linkers use passive actin network as a medium through which they  interact and self-organize.  After the symmetry of the  static  configuration is spontaneously broken the resultant active motion inside the cell produces overall steady translocation of the cell body.

The idea that contraction causes flow which in turn carries the regulators of contraction is incorporated into the hydrodynamic description of active fluids \cite{Kruse2003}. In static conditions, it has been shown to describe peaks in concentration of stress activator amplified by advective influx due to active stresses \cite{Bois2011}. In \cite{Hawkins2009a} similar idea  was used to describe initiation of non-lamellipodial motility associated with angular cortex flows. Heuristic models of the Keller-Segel type \cite{Perthame2008} describing polarization instability in  static cells with fixed length were proposed in \cite{Kruse2003a,Calvez2010}. In most of these models, however, the effect of contraction is obscured by the account of other mechanisms, in particular,  treadmilling, and the focus is on generation of internal flow rather than on the motion of a center of mass. There also exists considerable literature addressing spontaneous motility driven directly by protrusion \cite{John2008} and  Turing patterning \cite{Altschuler2008}, or studying interaction of multiple mechanisms \cite{Keren2008}.

To make the physics of mechanotaxis more transparent we study in this Letter the simplest analytically tractable 1D model which captures both the symmetry breaking and the induced  macroscopic  motion. To decouple dynamics of actin and myosin we assume infinite compressibility of the cytoskeleton \cite{Julicher2007}. In addition to active contractility the model accounts for long range elastic stiffness linking the front and the back of the cell \cite{Barnhart2010}.  

We show that initiation of motility in the mechanotaxis model is controlled by the average concentration of motor proteins.  The increase of motor concentration beyond a particular threshold leads to a bifurcation from a static symmetric regime to an asymmetric traveling wave (TW) regime describing a moving cell. While several TW regimes may be available for the same value of parameters,  stable  TW  solutions localize motors in the trailing edge of the cell in agreement with observations \cite{Verkhovsky1999}.

\emph{The model.} Consider the force balance equation for a 1D layer of an active gel in viscous contact with rigid background
 $\partial_{x}\sigma =\xi v $,
where $\sigma (x,t)$ is the stress, $v (x,t)$ is the velocity  and $\xi$ is the friction coefficient. Following \cite{Kruse2006,Julicher2007,Bois2011} we write
 $\sigma=\eta \partial_{x}v+\chi c$, where $\eta$ is the bulk viscosity, $c$ is the concentration of motors and  $\chi>0$ is the contractile pre-stress (per motor). The function $c(x,t)$ satisfies advection-diffusion equation $\partial_t c+\partial_x(c v)=D\partial_{xx}c,$ where $D$ is the diffusion coefficient.  We assume that $l_-(t)$ and $l_+(t)$ are the unknown boundaries of the cell. We also account for a mean field type linear elastic interaction due to membrane or cortex \cite{Barnhart2010} by using the following mechanical boundary condition, $\sigma(l_{\pm}(t),t) =-k(L-L_0)/L_0,$ where $L(t)=l_{+}(t)-l_{-}(t)$ is the length of the cell, $k$ is the effective elastic stiffness and $L_0$ is the reference length. Since we neglect  treadmilling we can write the  kinematic boundary conditions in the form
 $\dot{l}_{\pm}=v(l_{\pm}).$ Finally, we impose zero exterior flux of motors
 $\partial_{x}c (l_{\pm}(t),t)=0$ which implies that the average concentration $c_0=L_0^{-1}\int_{l_-}^{l_+}c(x,t)dx$ is conserved.

If we now normalize length by $L_0$, time, by $L_0^2/D$ and stress by $k$, we obtain a Keller-Segel type system
\begin{equation}\label{eq.1}
\begin{array}{c}
-\mathcal{Z}\partial_{xx}\sigma+\sigma=\mathcal{P}c/c_0,\\
\partial_t c+\mathcal{K}\partial_x(c \partial_x\sigma)=\partial_{xx}c,
\end{array}
\end{equation}
where the dimensionless constants are $\mathcal{Z}= \eta/(\xi L_0^2)$,
 $\mathcal{K}=k/(\xi D)$ and $\mathcal{P}=c_0 \chi / k$. If $\sigma$ is expressed through the corresponding Green's function, the resulting nonlocal diffusion-advection problem is structurally similar to the one proposed in \cite{Kruse2003a}, however the effective kernel is different.

The dimensionless boundary conditions for (\ref{eq.1}) take the form
$\sigma(l_{\pm}(t),t)=-(L(t)-1)$,
$\partial_{x}c (l_{\pm}(t),t)=0$ and
$\dot{l}_{\pm}(t)=\mathcal{K} \partial_x\sigma(l_{\pm}(t),t)$. They imply that the motion of the center of the cell $G(t)=(l_-(t)+l_+(t))/2$ is governed by the equation:
\begin{equation}\label{ODE}
\dot{G}(t)=\frac{\mathcal{KP}}{2\mathcal{Z}c_0}\int_{l_{-}(t)}^{l_{+}(t)}\frac{\text{sh}((G-x)/\sqrt{\mathcal{Z}})}{\text{sh}(L/(2\sqrt{\mathcal{Z}}))} c(x,t)dx.
\end{equation}
One can see that if the concentration distribution is symmetric then $\dot{G}=0$ and the cell cannot move, which is a simple analog of Purcell's theorem \cite{Purcell1977} with spatial asymmetry replacing temporal asymmetry. From (\ref{ODE})  one can also infer that the maximal speed of the cell is equal to $\mathcal{KP}/(2\mathcal{Z})$. In dimensional variables \cite{Julicher2007, Bois2011} this gives  $\chi L_0 c_0/(2\eta)\simeq 10\mu m/min $ which is realistic \cite{Verkhovsky1999}.

\emph{TW regimes.} To study the traveling wave regimes we assume that both stress and myosin concentration depend on the moving coordinate $y=x-Vt$ where $V$ is the unknown cell velocity. We also put $\dot{l}_{\pm}=V$  and $L(t)=L$ where $L$ is the unknown length of the cell. System (\ref{eq.1}) reduces to a single equation
\begin{equation} \label{eq.2}
- \mathcal{Z}s^{''}+ s - \mathcal{K}(L-1)= \mathcal{KP} \frac{\exp(s-Vy)}{\int_0^L \exp(s-Vy)dy}
\end{equation}
where $s(y) = \mathcal{K}\left[ \sigma(y)+(L-1)\right]$ is the unknown function. The presence of four boundary conditions, $s(0)=s(L)=0$ and  $s'(0)=s'(L)=V$, ensures that both parameters $V$ and $L$ can be found along with $s(y)$. After equation (\ref{eq.2}) is solved the motor concentration profile can be recovered from a relation $ c(y)= c_0  \exp(s(y)-Vy)/[\int_0^L\exp(s(y)-Vy)dy].$ To simplify the description we first assume that $\mathcal{Z}=1$ \cite{Julicher2007} which means that the elastic and the viscous scales in a cell are correlated. We are then left with two dimensionless parameters $\mathcal{K}\sim 100$ and $\mathcal{P}\sim 0.1$ \cite{Julicher2007,Hawkins2009a}, where $\mathcal{K}$ is the measure of internal stiffness while $\mathcal{P}$ gives the scale of motor activity.
\begin{figure}[!h]
\begin{center}
\includegraphics[scale=0.22]{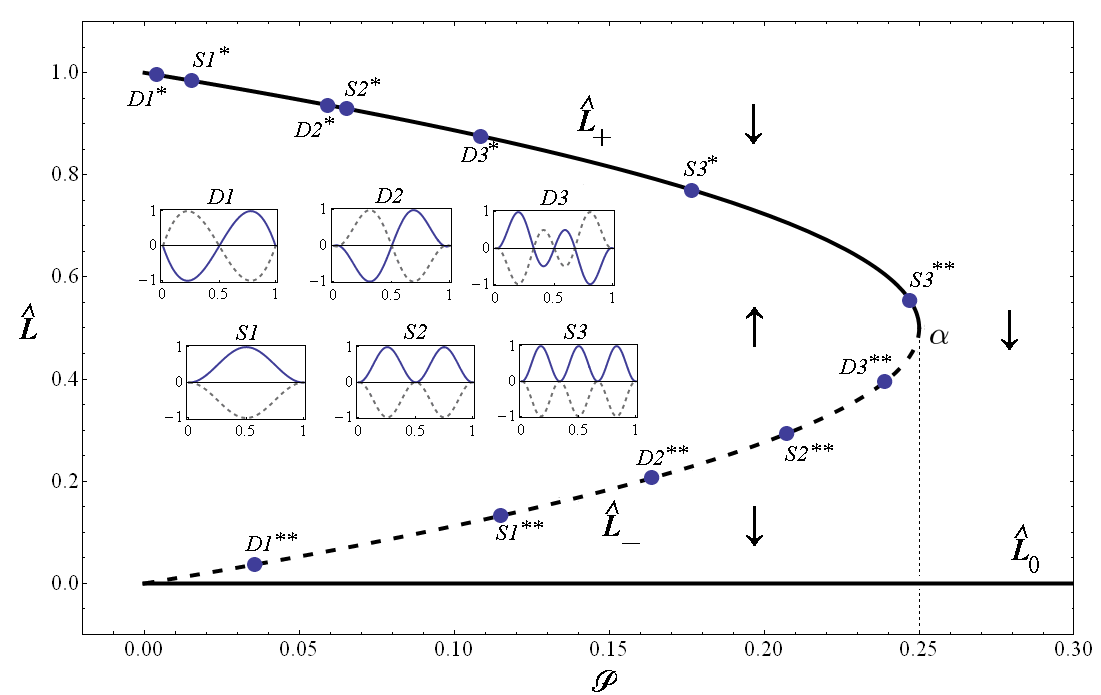}
\caption{\label{bifur2}  Three families of static solutions ($\hat{L}_+, \hat{L}_-$ and $\hat{L}_0$ parameterized by $\mathcal{P}$ at $\mathcal{K}=2600$. The bifurcation points are labeled as $D1,D2,...$, when the nontrivial bifurcated solution is  motile  ($\delta V \neq 0$) and by $S1,S2,...$ when it is static ($\delta V = 0$). Inserts show the eigenfunctions $\delta s(y/L)$. In the inserts solid and dashed lines distinguish eigenfunctions with positive and negative amplitudes ($\delta L$ or $\delta V$).}
\end{center}
\end{figure}

The initiation of motility is associated with an instability of a  static  solution of (\ref{eq.2}) with $V=0$. All such solutions can be written in quadratures and all of them except the homogeneous ones imply internal flow  \cite{Bois2011}. In addition to the regular  static  solutions there are also  singular  static  solutions with zero length $\hat{L}_0=0$ and $s(y)=\lim_{\theta\rightarrow 0} \theta f(y/\theta),$  where $f(u)=(\mathcal{KP}/2) u(1-u)$ and $u\in [0,1]$; moreover, for $\mathcal{P}>1/4$, those are the only  static  configurations. Measure valued solutions of this type are known in related fields \cite{singular} and here they describe the collapsed cells under the action of unbalanced contractile stresses.

\begin{figure}[!h]
\begin{center}
\includegraphics[scale=0.25]{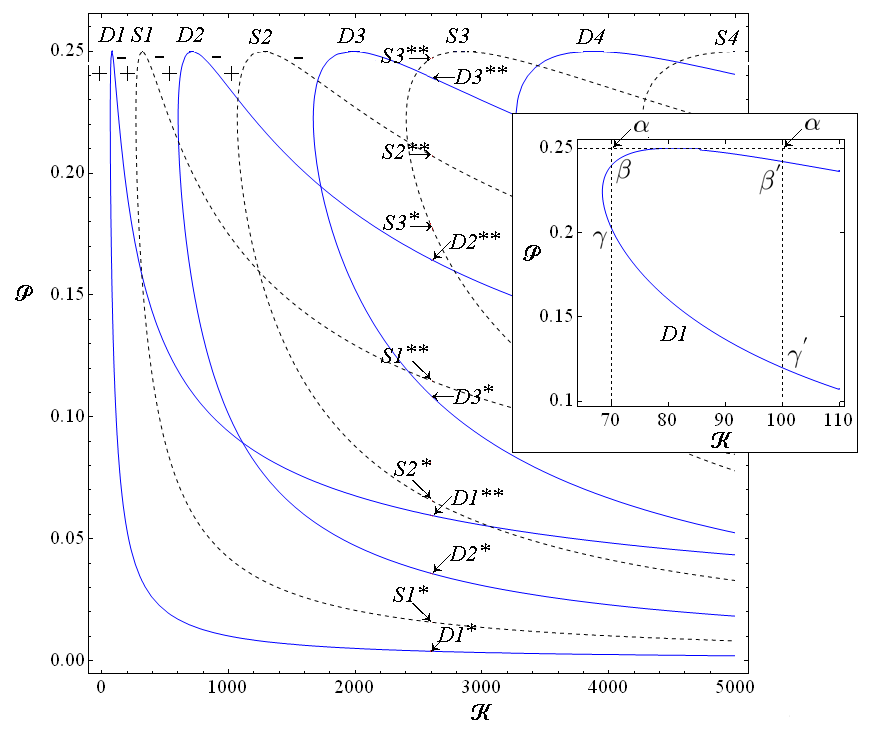}
\caption{\label{bifur1} Locus of the bifurcation points in the $(\mathcal{K},\mathcal{P})$ plane. Insert shows a zoom on the $D1$ branch around the turning point at $\mathcal{P}=1/4$.  The detailed bifurcation diagrams for $\mathcal{K}=2600$   and $\mathcal{P}=0.245$ are shown in Fig.\ref{bifur2} and Fig.\ref{bifur_K} from where the meaning of labels $\beta$, $\gamma$, $\beta^{'}$ $\gamma^{'}$ becomes clear.  }
\end{center}
\end{figure}

To show that  motile  branches with $V\neq 0$ can bifurcate only from homogeneous  static  solutions with $s(y)=0$, $V=0$ and
\begin{equation}\label{nonmotileL}
\hat{L}_{\pm}=(1\pm\sqrt{1-4 \mathcal{P}})/2,
\end{equation}
we observe that for $V\neq 0$ equation (\ref{eq.2}) has an integral $ L^{-1}\int_0^L\exp(s(y)-Vy)dy=(1-\exp(-LV))/(LV)$ which in the limit $V\rightarrow 0$ gives $\int_0^{L}\exp(s)=L$. Since in static solutions $s(y)$ must necessarily have a constant sign, this integral implies that $s(y)=0$ and hence such solutions must be trivial. As we show in Fig.\ref{bifur2} there are two families of non-singular trivial solutions: with longer ($\hat{L}_{+}$ family)  and shorter ($\hat{L}_{-}$ family) lengths.

Linearization around these trivial solutions produces the following linear problem for the function $\delta s(y/L)$
\begin{equation}\label{eq.21}
\delta s^{''}+\omega^2\delta s=A+By/\hat{L}
\end{equation}
where $ \omega =\sqrt{\mathcal{K}\mathcal{P}\hat{L}-\hat{L}^2}$, $A=-( \omega^2+\hat{L}^2)[(2 \hat{L}-1)/( \hat{L}^3(\hat{L} -1))\delta L-(1/2)\hat{L} \delta V]$ and $B=-\hat{L}( \omega^2+ \hat{L}^2)\delta V.$ Equation (\ref{eq.21}) is supplemented with four  boundary conditions $\delta s (0)=\delta s (1)=0$, $\delta s ^{'}(0)=\delta s ^{'}(1)= \hat{L} \delta V$ allowing one to find the parameters $\delta L$ and $\delta V$ (up to a multiplier). Problem (\ref{eq.21}) has  nontivial solutions if $ \omega\neq 0$ and
\begin{equation}\label{bifurc_val}
2 \hat{L}^2(\cos \omega-1)+(\omega^2+\hat{L}^2) \omega\sin\omega=0.
\end{equation}

\begin{figure}[!h]
\begin{center}
\includegraphics[scale=0.25]{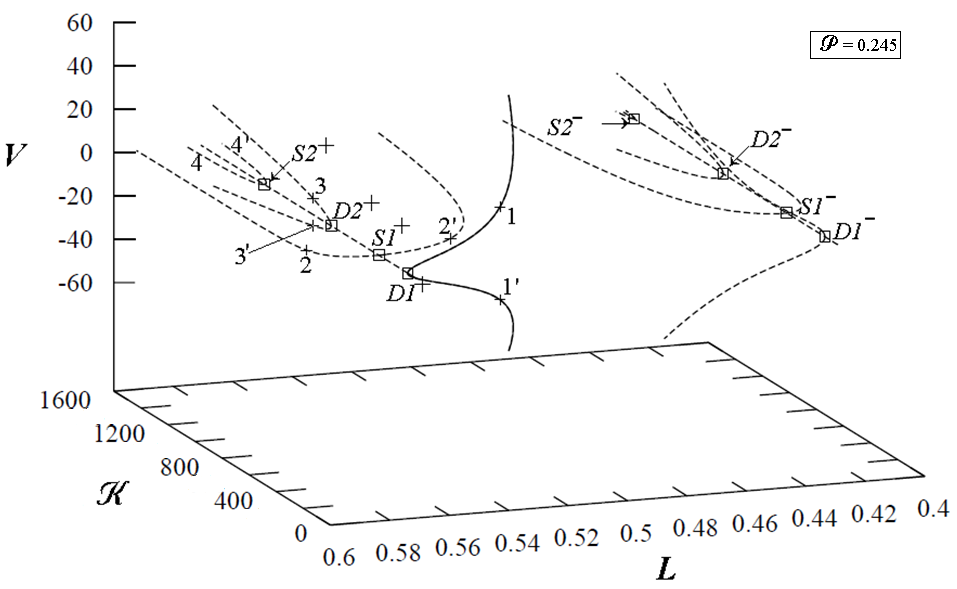}
\caption{\label{bifur_K} Bifurcation diagram with $\mathcal{K}$ as a parameter showing nontrivial solutions branching from families of homogeneous static solutions $\hat{L}_{+}$  and $\hat{L}_{-}$. The value $\mathcal{P}=0.245$ is fixed. Solid lines show stable motile branches while all the dotted lines correspond to unstable solutions. The internal configurations corresponding to branches indicated by numbers $(1,1',2,2',etc)$ are shown in Fig.\ref{profiles}.  }
\end{center}
\end{figure}

\begin{figure}[!h]
\begin{center}
\includegraphics[scale=0.3]{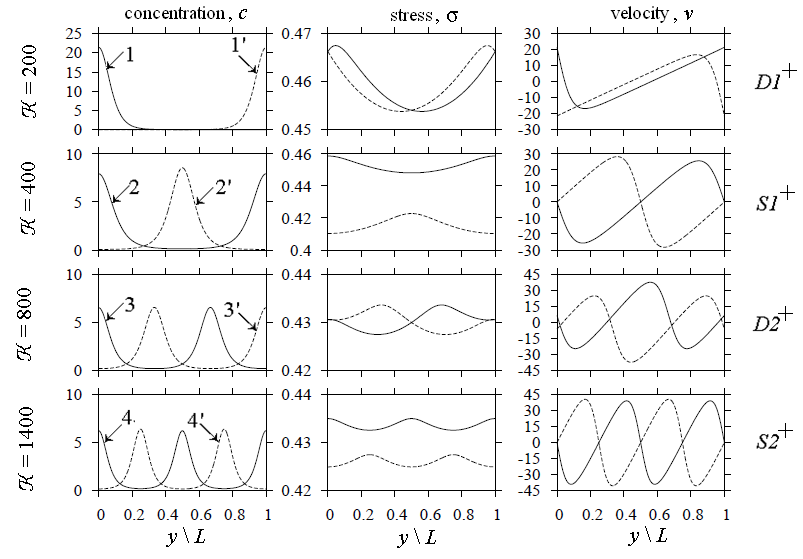}
\caption{\label{profiles} Internal profiles associated with successive bifurcated solutions shown in Fig.\ref{bifur_K} for $\mathcal{P}=0.245$: (1,3) correspond to asymmetric motile branches while (2,4) describe symmetric  static  branches. }
\end{center}
\end{figure}

Solutions of the characteristic equation (\ref{bifurc_val}) can be split in two families. The first family, $\omega =2m\pi$ with $m$ a positive integer, corresponds to static configurations with  $\delta V=0$ and $\delta s(y/L))=\delta L (1-\cos(\omega y/\hat{L}))$. On the parameter plane ($\mathcal{P},\mathcal{K}$), see Fig.\ref{bifur1}, sub-families of bifurcational points corresponding to different $m$ will be labeled as $S1^{\pm},S2^{\pm},...$ at constant $\mathcal{P}$
where superscripts $\pm$ indicate branches $ \hat{L}_{\pm}$ from (\ref{nonmotileL}). It will be also convenient to distinguish as $S1^{*}/S1^{**}, etc. $  bifurcational points corresponding to longer/shorter  static  configurations  at fixed $\mathcal{K}$.  The second family, defined by the equation $\tan(\omega/2 )=(\omega/2)(1+\omega^2/\hat{L}^2)$ and indexed as $D1^{\pm},D2^{\pm},...$ corresponds to  motile  solutions with  $\delta L=0$,  $\delta s(y/L)=\delta V (\sin(\omega(y/\hat{L}-1/2))-\sin(\omega/2)(2(y/\hat{L})-1))$;   the notations  $D1^{*}/D1^{**}$ will have the same meaning as in the case of static solutions. The locus of the bifurcation points in the parameter plane ($\mathcal{P},\mathcal{K}$)  is shown in Fig.\ref{bifur1}. Each branch (say, $D1$) is represented by two segments ($D1^+$ and $D1^{-}$) that meet smoothly  at $\mathcal{P}=1/4$.

To follow the bifurcated branches into the nonlinear regime we  performed a numerical study of the equation (\ref{eq.2}). A bifurcational diagram at fixed $\mathcal{P}$, showing both static and motile configurations is presented in Fig.\ref{bifur_K}; the corresponding internal profiles are shown in Fig.\ref{profiles}.
\begin{figure}[!h]
\begin{center}
\includegraphics[scale=0.3]{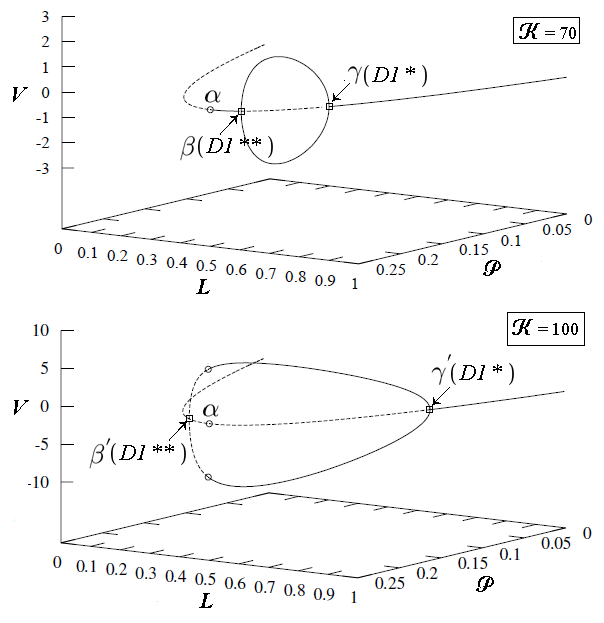}
\caption{\label{bifur_P} Bifurcation diagram with $\mathcal{P}$ as a parameter showing motile branches connecting points $D1^*$ and  $D1^{**}$. Corresponding bifurcation points are shown in insert in Fig.\ref{bifur1}. Parameter $\mathcal{K}$ is fixed in each graph $(\mathcal{K}=70 \text{ and } \mathcal{K}=100)$.}
\end{center}
\end{figure}
We see that each of these pitchfork bifurcations gives rise to two nontrivial solutions. For instance, point $D1^+$  is associated with two motile  branches,  the point  $S1^+$ - with two static branches.  Each pair of  motile  solutions is symmetric with two opposite polarization orientations corresponding to two different signs of the velocity. Along the first motile branch originating at $D1^+$   motors  always concentrate at the trailing edge.  For the second  motile  branch originating at $D2^+$ there is an additional peak in the concentration profile (Fig.\ref{profiles}).  The static bifurcation point $S1^+$ gives rise to two symmetric configurations with different lengths and with motors concentrated either in the middle of the cell or near the boundaries (Fig.\ref{profiles}). The higher order static and   motile  bifurcation points produce solutions with more complex internal patterns. For the branches bifurcating from the trivial configurations belonging to $\hat{L}_{-}$ family, the picture is similar (see Fig.\ref{bifur_K}).
\begin{figure}[!h]
\begin{center}
\includegraphics[scale=0.3]{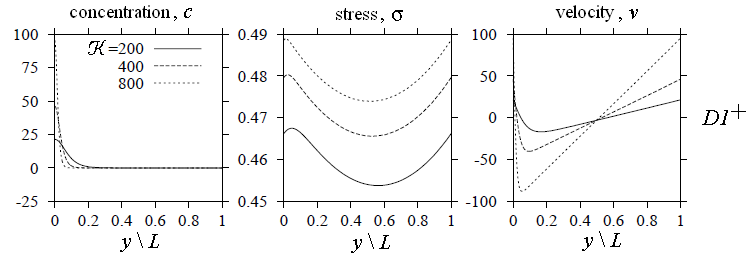}
\caption{\label{localization} Internal configuration of the moving cell on the motile branch $D1^+$ showing the localization with increasing $\mathcal{K}$ at $\mathcal{P}=0.245$. }
\end{center}
\end{figure}

In Fig.\ref{bifur_P} we show in more detail the nontrivial solutions originating from the  motile  branch $D1$ at two values of parameter $\mathcal{K}$ corresponding to lines $\alpha \beta$ and $\alpha\beta^{'}$ shown in Fig.\ref{bifur1} (insert). One can see that there is a single solution connecting points $D1^*$ and $D1^{**}$ which may belong either to one family  $\hat{L}_{+}$  ($\alpha \beta$) or to two different families $\hat{L}_{+}$  and $\hat{L}_{-}$ ($\alpha\beta^{'}$). In the former case the nontrivial  motile  branch has a turning point at a finite value of  $\mathcal{P}<1/4$ giving rise to a  reentrant behavior first observed in \cite{Kruse2003a}. In this regime the increase of the average concentration of  myosin first polarizes the cell and initiates motility, but then if the concentration is increased further, the cell get symmetrized again and stabilize in another static homogeneous configuration.

\emph{Non-steady transients.} A study of the initial value problem (\ref{eq.1}) shows that all nontrivial solutions (static  and  motile)  are unstable except  for the branch bifurcating at $D1^+$. Homogeneous solutions from the $\hat{L}_{+}$ family and all singular static solutions  from the $\hat{L}_{0}$ family are stable. Numerical simulations also suggest that as in \cite{Bois2011, Kruse2003a}, unstable multi-peaked solutions are long living. This behavior is reminiscent  of  the classical spinodal decomposition modeled by 1D Cahn-Hilliard equation where the coarsening process get critically slowed down near multiple saddle points \cite{Carr89}.

In some limiting cases the mechanotaxis equations can be  simplified but the solutions become more singular. Thus, in the hyperbolic limit $\mathcal{K}\rightarrow \infty$ (no diffusion), the number of nontrivial solutions grows to infinity while the solutions become measure valued.  For instance,  as we show in  Fig.\ref{localization}, the concentration profile  for the first motile branch  ($D1^+$) infinitely localizes at the trailing edge.  In the inviscid limit $\mathcal{Z}\rightarrow 0$ the system (\ref{eq.1}) reduces to $\partial_t u= \partial_x(u\partial_x u)$, where $u = 1-\mathcal{KP}c/c_0$, which is a sign-indefinite porous flow equation exhibiting an uphill diffusion when $c/c_0>(\mathcal{KP})^{-1}$. If the cell length is fixed meaning $k \rightarrow \infty$,  we have $\mathcal{K}\rightarrow\infty$ and $\mathcal{P}\rightarrow 0$ and it is more convenient to restore $\mathcal{Z}$ and use as a dimensionless parameter the surviving product  $\mathcal{KP}$ which is proportional to the contraction-based Peclet number $\chi/(D \xi)$ introduced in \cite{Bois2011, Hawkins2009a}; in this case all inhomogeneous solutions are static and can be described in quadratures.

In conclusion, we proposed a prototypical model of a crawling cell showing the possibility of spontaneous polarization leading to steady self propulsion in the conditions when contraction is the only active process while treadmilling is disabled. This model complements the existing theories of polarization which place emphasis on treadmilling. The model reduces to a Keller-Segel type system, however, here the nonlocality is due to mechanical rather than chemical feedback. We obtained a variety of  motile  TW regimes corresponding to finite size self propelling active bodies with free boundaries. Similar to the Navier-Stokes system, where nonlocality is hidden behind the incompressibility assumption, the  system of mechanotaxis equations has quadratic nonlinearity and shows an infinite sequence of bifurcations as the diffusion coefficient goes to zero.

\bibliographystyle{tPHM}

\end{document}